\documentclass[onecolumn,showpacs,preprintnumbers,amsmath,amssymb]{revtex4}

\usepackage{graphicx}
\usepackage{epsfig}

\def\dd{\displaystyle}

\begin{document}
\title{\bf The Casimir Energy For Scalar Field With Mixed Boundary Condition}
\author{M. A. Valuyan}
\email{m-valuyan@sbu.ac.ir; m.valuyan@semnaniau.ac.ir}
\affiliation{Department of Physics, Semnan Branch, Islamic Azad University, Semnan, Iran}
\date{\today}

\begin{abstract}
In the present study, the first-order radiative correction to the Casimir energy for massive and massless scalar fields confined with mixed boundary conditions\,(Dirichlet-Neumann) between two points in $\phi^4$ theory was computed. Two issues in performing the calculations in this work are essential: to renormalize the bare parameters of the problem, a systematic method was employed, allowing all influences from the boundary conditions to be imported in all elements of the renormalization program. This idea yields our counterterms appeared in the renormalization program to be position-dependent. Using the Box Subtraction Scheme\,(BSS) as a regularization technique is the other noteworthy point in the calculation. In this scheme, by subtracting the vacuum energies of two similar configurations from each other, regularizing divergent expressions and their removal process were significantly facilitated. All the obtained answers for the Casimir energy with the mixed boundary condition were consistent with well-known physical grounds. We also compared the Casimir energy for massive scalar field confined with four types of boundary conditions\,(Dirichlet, Neumann, mixed of them and Periodic) in $1+1$  dimensions with each other, and the sign and magnitude of their values were discussed.
\end{abstract}

\maketitle

\section{Introduction}
\label{sec:intro}
The study of Boundary Conditions\,(BCs) in the theory of quantum fields is highly broad and almost impossible to ignore the effect of the BCs on this topic. Whereas, the Casimir energy and its related force were defined as the difference between two vacuum energies in presence and absence of BCs, this topic was found unavoidably related to all sorts of BCs imposed on quantum fields\,\cite{h.b.g.}. In a vast array of previous papers, the Casimir energy was discussed in multiple geometries with different BCs\,\cite{review.1,Neumann.1,Int.Geo.1,Neumann.2,phys.rep.,Int.Geo.2,BSS.Curved}. Bordag et al. conducted the first study on the Casimir energy for self-interacting quantum field theory\,\cite{Bordag.et.al.}, and later it was also found an increasing attention for multiple BCs in different geometries\,\cite{other.RC.2}. What is commonly the source of challenge in radiative correction to the Casimir energy is how to perform a proper renormalization program in which the bare parameters of the problem would be completely renormalized. In many previous works, in performing the renormalization program even for bounded fields, to renormalize the bare parameters, \emph{free counterterms} have been used\,\cite{free.counterterms.1,free.counterterms.2,free.counterterms.3,free.counterterms.4,NN-DN-DD}. Our meaning for the \emph{free counterterms} is the one used in unbounded spaces\,(free space). We maintain that, the use of free counterterms in all problems, regardless of the type of the dominant BCs on the field cannot be correct. Therefore, it is necessary that a renormalization program be performed to consider the effects of the BC on this process. This idea has been addressed in several previous works; however, the renormalization program proposed by Gousheh et al. is highly a simple and systematic program in renormalizing the bare parameters in bounded spaces\,\cite{1D-Reza,3D-Reza}. In their proposed program, the counterterms responsible for eliminating divergences due to the bare parameters of the lagrangian were obtained position-dependent. In fact, it can be interpreted that obtaining a position-dependent counterterms reflects the influence of BCs on the renormalization program. Later, this type of renormalization program has been used in multiple geometries defined in the flat and curved manifolds, and the achieved results for each case were physically consistent\,\cite{2D-Man,other.RC.1}. In this study, maintaining their idea and for the first time, we calculate the radiative correction to the Casimir energy between two points confined with the mixed BC in $1+1$ dimensions. The final result for this quantity was also consistent with the known physical basis and the renormalization procedure was conducted as simple as possible.
\par
\begin{figure} \hspace{0cm}\includegraphics[width=8cm]{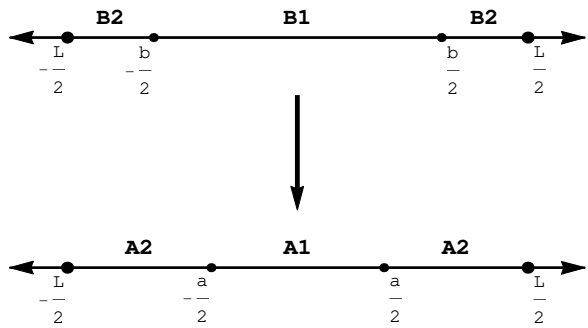}
\caption{\label{Fig.1} \small
  The geometry of the two different configurations whose energies are to be compared. The
  labels $A1$, $B1$, etc. denote the appropriate sections in each configuration separated by points. The
  upper configuration is denoted by $\mathcal{B}$, and the lower one by $\mathcal{A}$.}
\end{figure}

The ongoing conflict between the computation process of the Casimir energy and the divergences has made the use of the regularization techniques an unavoidable part of these sorts of computations. In this study, to regularize the appeared infinities in the calculation procedures, we used the Box Subtraction Scheme\,(BSS) as a regularization technique. In this scheme, two similar configurations are usually introduced and to achieve the Casimir energy, the vacuum energy of these two configurations is subtracted from each other. According to Fig.\,(\ref{Fig.1}), we define two similar configurations. In the lower picture in Fig.\,(\ref{Fig.1}), the original system\,(two points with distance $a$) is trapped in a box\,(\emph{e.g.}, two points with distance $L$). This configuration was named as ``$\mathcal{A}$'' configuration. Analogous to the configuration ``$\mathcal{A}$'', we defined the ``$\mathcal{B}$'' configuration in which two points with distance $b$ were trapped in the other similar box\,(\emph{i.e.}, two points with distance $L>b$). The labels $A1$, $A2$, $B1$ and $B2$ denote the sections in each configuration separated by points. To obtain the Casimir energy between two points with distance $a$, the following definition should be introduced:
\begin{equation}\label{BSS.Def.}
  E_{\hbox{\tiny{Cas.}}}=\lim_{L\to\infty}\lim_{b\to\infty}[E_\mathcal{A}-E_\mathcal{B}],
\end{equation}
where $E_\mathcal{A}$ and $E_\mathcal{B}$ are the vacuum energies of configurations $\mathcal{A}$ and $\mathcal{B}$, respectively. This scheme was based on the method used by Boyer\,\cite{Boyer.} and the significant decrease in the need to use analytic continuation technique was reported as the main advantage of this scheme. We can interpret that subtraction of the vacuum energy of the configuration ``$\mathcal{B}$'' from the ones for the configuration ``$\mathcal{A}$'' is the work conducted in deforming the configurations ``$\mathcal{B}$'' to ``$\mathcal{A}$''. This interpretation can guarantee that the final result of the subtraction is finite. This advantage is helpful, especially in the even spatial dimensions, which the Casimir energy in this dimension leads to a divergent answer\,\cite{2D-Man,cavalcanti.}. Moreover, the BSS presents a simple and clear method in removing the divergences from the Casimir energy calculations. In the usual definition of the Casimir energy, the vacuum energy of the free space\,(Minkowski space) is subtracted from the vacuum energy of the bounded space. However, in the BSS, the Minkowski space was replaced by a secondary configuration\,(\emph{e.g.}, the configuration ``$\mathcal{B}$'') in proper limits. This type of definition for the Casimir energy imports more parameters in the calculation procedures. These additional parameters play the role of regulators in the computation process, resulting in performing the removal process of the divergences with more clarity.
\par
\begin{figure}
    \hspace{0cm}\includegraphics[width=8.5cm]{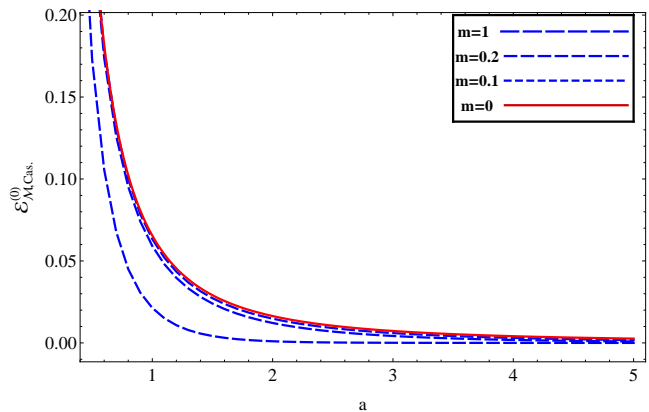}
 \caption{\label{ZMMasslessCH}   The zero-order Casimir energy density for massive and massless scalar fields are plotted as a function of the distance
between the points ($a$), within the free scalar field theory. In this figure we have shown the sequence of plots for $m=\{1,0.2,0.1,0\}$. It is apparent that the sequence of the massive cases converges rapidly to the massless case. }
\end{figure}

The leading order of the Casimir energy for massive scalar field confined between two parallel plates with mixed BC\,(Dirichlet-Neumann) in arbitrary $d$ spatial dimensions has already been computed previously\,\cite{NN-DN-DD,wolfram.}. Hence, in this study, we have only reported the final answer of this quantity in $1+1$ dimensions and the calculation procedure for this order of energy is not repeated. Accordingly, the Casimir energy expression for massive scalar field confined between two points with mixed BC in zero order of $\lambda$ for $\phi^4$ theory is:
\begin{equation}\label{MCas.Massive}
  E^{(0)}_{\hbox{\tiny{$\mathcal{M}$,Cas.}}}=\frac{m}{2\pi}\sum_{j=1}^{\infty}\frac{(-1)^{j+1}K_1(2maj)}{j}.
\end{equation}
The two particular limits, which are usually discussed in the Casimir energy, are the massless and large-mass limits. For these two limits, the above equation converts to:
\begin{equation}\label{massless-largemass-ZeroMixed}
   E^{(0)}_{\hbox{\tiny{$\mathcal{M}$,Cas.}}}\longrightarrow\left\{
                                       \begin{array}{ll}
                                         \dd\frac{\pi}{48a}, &\hspace{1.5cm} \hbox{$m\to0$;} \\ \\
                                        \dd \frac{1}{4}\sqrt{\frac{m}{\pi a}}e^{-2ma}, & \hspace{1.5cm}\hbox{$ma\gg1$.}
                                       \end{array}
                                     \right.
\end{equation}
In Fig.\,(\ref{ZMMasslessCH}), the Casimir energy density for the massive and massless scalar field as a function of the distance of the points ($a$) were plotted. This figure shows that the Casimir energy density value for the massive cases approaches to one for the massless case when the mass of the field goes to zero.
\par
In the next section, the first-order radiative correction to the Casimir energy for the massive and massless scalar fields with the mixed BC\,(Dirichlet-Neumann) in $1+1$ dimensions between two points was computed. Use of the aforementioned renormalization program supplemented by the BSS as a regularization technique leads more clarity in this calculation process. Finally, we will outline all the obtained results of the Casimir energy with four Dirichlet, Neumann, Periodic and mixed BCs.

\section{Radiative Correction to The Casimir Energy}
\label{sec:RC.Cas.Cal.}
The lagrangian with the quadratic self-interacting term for real massive scalar field in $\phi^4$ theory can be written as:
\begin{equation}\label{lagrangian}
  \mathcal{L}= \frac{1}{2}[\partial_\mu \phi(x)]^2-\frac{1}{2}m_0^2\phi(x)^2-\frac{\lambda_0}{4!}\phi(x)^4,
\end{equation}
where $m_0$ and $\lambda_0$ are the bare mass of the field and bare coupling constant, respectively. To renormalize the bare parameters of the lagrangian, a renormalization program should be conducted. In the literature, the vast variety of discussions on the proper renormalization program were reported. As stated in the Introduction, Gousheh et al. proposed a systematic program allowing for importing the effects of the BCs in the renormalization program \,\cite{1D-Reza}. The counterterms that they used in the renormalization program, are position-dependent due to taking the consistency with the imposed BC. Details of the calculation to obtain the proper counterterms and attain the expression for the vacuum energy up to first-order of coupling constant $\lambda$  were discussed previously, and in the following, by using their proposed program, we have\,\cite{1D-Reza}:
\begin{equation}\label{first.order.vacuum.energy.}
  E^{(1)}=\frac{-\lambda}{8}\int G^2(x,x)dx,
\end{equation}
where $E^{(1)}$ denotes the total vacuum energy expression up to first order of $\lambda$, and $G(x,x)$ is the Green's function. In the mixed BC, the Dirichlet and Neumann BCs are used simultaneously. Precisely, to apply this type of BC on two parallel plates, the Dirichlet BC should be satisfied on the left plate (\emph{e.g.}, the plate placed on $x=\frac{-a}{2}$) and on the opposite side, the plate placed on the right\,($x=\frac{a}{2}$) satisfies the Neumann BC. It should be noted that for the reverse order in applying the Dirichlet and Neumann BCs on the plates, the vacuum energy expression remains unchanged. For the original region $A1$ displayed in Fig.(\ref{Fig.1}), after performing the usual computations in determining of the Green's function expression, this expression for the massive scalar field confined with mixed BC between two points located on $x=\frac{\pm a}{2}$ in $1+1$ dimensions after Weak rotation becomes:
\begin{equation}\label{Green.Mixed.}
  G_{\hbox{\tiny{$\mathcal{M}$,$A1$}}}(x,t,x',t')=\frac{1}{a}\int\frac{d\omega}{2\pi}\sum_{n=0}^{\infty}\frac{e^{-\omega(t-t')}[(-1)^n\cos(k_n x)+\sin(k_n x)][(-1)^n\cos(k_n x')+\sin(k_n x')]}{\omega^2+\omega_n^2},
\end{equation}
where $k_n=\frac{(2n+1)\pi}{2a}$ and $\omega^2_n=k_n^2+m^2$ are the wave-vector and wave-number, respectively. By substituting the Green's function expression presented in Eq.\,\eqref{Green.Mixed.} with Eq.\,\eqref{first.order.vacuum.energy.}, the vacuum energy for region $A1$ displayed in Fig.\,(\ref{Fig.1}) is obtained as:
\begin{equation}\label{Vacuum.A1.Mixed.RC}
  E^{(1)}_{\hbox{\tiny{$\mathcal{M}$,$A1$}}}=\frac{-\lambda}{32a}\sum_{n=0}^{\infty}\sum_{n'=0}^{\infty}
  \frac{1}{\omega_n\omega_{n'}}\left(1+\frac{1}{2}\delta_{n,n'}\right).
\end{equation}
The first term of above equation is divergent, and to regularize its infinity, we should first convert the summation into the integral form. For this purpose, the following form of Abel-Plana Summation Formula\,(APSF) written for half integer parameters is used\,\cite{APSF}:
\begin{equation}\label{APSF}
  \sum_{n=0}^{\infty}\mathcal{F}(n+\frac{1}{2})=\int_{0}^{\infty}\mathcal{F}(x)dx-i\int_{0}^{\infty}\frac{\mathcal{F}(it)-\mathcal{F}(-it)}{e^{2\pi t}+1}dt.
\end{equation}
The first term on the right hand side\,(rhs) of the above equation is usually divergent, and we have henceforth named it as \emph{integral term}. The last term on the rhs of the above equation is usually convergent, and we have named it as \emph{Branch-cut} term. Therefore, we have:
\begin{eqnarray}
\label{rc.after.APSF.Mixed}
  E^{(1)}_{\hbox{\tiny{$\mathcal{M}$,$A1$}}}&=&\frac{-\lambda}{32a}\left[\left(\sum_{n=0}^{\infty}\frac{1}{\omega_n}\right)^2+\frac{1}{2}
  \sum_{n=0}^{\infty}\frac{1}{\omega^2_n}\right]\nonumber\\
  &=&\frac{-\lambda}{32a}\bigg\{\bigg[\underbrace{\frac{a}{\pi}\int_{0}^{\infty}\frac{d\xi}{\sqrt{\xi^2+1}}}_{\mathcal{I}_\infty(a)}
  +B(a)\bigg]^2+\frac{a}{4m}\tanh\left(ma\right)\bigg\},
\end{eqnarray}
where $\mathcal{I}_\infty(a)$ is the integral term of APSF, and it has a divergent value. $B(a)$ is the Branch-cut term of APSF and its value is:
\begin{eqnarray}\label{Branch-cut. term.}
B(a)=\frac{-2a}{\pi}\int_{1}^{\infty}\frac{d\eta}{\sqrt{\eta^2-1}(e^{2ma\eta}+1)}
=\frac{2a}{\pi}\sum_{j=1}^{\infty}(-1)^{j}\int_{1}^{\infty}\frac{e^{-2ma\eta j}}{\sqrt{\eta^2-1}}d\eta
=\frac{2a}{\pi}\sum_{j=1}^{\infty}(-1)^{j}K_0(2maj).
\end{eqnarray}
Now, using Eq.\,\eqref{rc.after.APSF.Mixed} and the BSS defined in Eq.\,\eqref{BSS.Def.}, the subtraction of the vacuum energy of regions becomes:
\begin{eqnarray}
\label{subtraction.Mixed.rc}
  E_\mathcal{A}-E_\mathcal{B}=\frac{-\lambda}{32a}\left[\mathcal{I}^2_\infty(a)+2\mathcal{I}_\infty(a)B(a)+B^2(a)
  +\frac{a}{4m}\tanh\left(ma\right)\right]\hspace{6.7cm}\nonumber\\
  +2\frac{-\lambda}{32\frac{L-a}{2}}\left[\mathcal{I}^2_\infty\Big(\frac{L-a}{2}\Big)
  +2\mathcal{I}_\infty\Big(\frac{L-a}{2}\Big)B\Big(\frac{L-a}{2}\Big)+B^2\Big(\frac{L-a}{2}\Big)+\frac{L-a}{8m}\tanh\left(\frac{m(L-a)}{2}\right)\right]
  -\{a\to b\}.
\end{eqnarray}
For the first term in the brackets of the above equation, we have:
\begin{equation}\label{first.term.bracket.}
  \frac{-\lambda}{32}\left[\frac{1}{a}\mathcal{I}^2_\infty(a)+2\frac{2}{L-a}\mathcal{I}^2_\infty\Big(\frac{L-a}{2}\Big)-\{a\to b\}\right]
  =\frac{-\lambda}{32\pi^2}\left[a+2\frac{L-a}{2}-b-2\frac{L-b}{2}\right]\left(\int_{0}^{\infty}\frac{d\xi}{\sqrt{\xi^2+1}}\right)^2=0.
\end{equation}
As Eq.\,\eqref{first.term.bracket.} shows, all divergent terms are removed for any finite values of $a$, $b$ and $L$ and the most important role in this removal process is played by the BSS. The second term in the brackets of Eq.\,\eqref{subtraction.Mixed.rc} is still divergent, and to remove its infinity, we replace the upper limit of the integral $\mathcal{I}$ with a cutoff and calculate it and expand the result as follows:
\begin{equation}\label{I.calculation.}
  \mathcal{I}_\Lambda(a)=\frac{a}{\pi}\int_{0}^{\Lambda}\frac{d\xi}{\sqrt{\xi^2+1}}
       \buildrel \Lambda\to\infty\over{\longrightarrow}\frac{a}{\pi}\left(\ln2+\ln\Lambda+\mathcal{O}(\Lambda^{-2})\right).
\end{equation}
This expansion causes the infinite part of the integration to be manifested. Analogously, this scenario is conducted for the same terms in the other regions. It can be shown that proper adjusting for cutoffs causes all divergent parts related to $\mathcal{I}(x)$ to be removed via BSS. Therefore, for the second term in the brackets of Eq.\,\eqref{subtraction.Mixed.rc}, we have:
\begin{eqnarray}\label{Hazfeh-ILambda}
  \frac{-\lambda}{16\pi}\left[\left[\ln2+\ln\Lambda+\mathcal{O}(\Lambda^{-2})\right]B(a)
   +2\left[\ln2+\ln\Lambda'+\mathcal{O}(\Lambda'^{-2})\right]B\big(\frac{L-a}{2}\big)-\{a\to b\}\right]\hspace{3.1cm}\nonumber\\
    \buildrel{\hbox{\tiny{BSS}}}\over\longrightarrow \frac{-\lambda}{16}\left[\frac{1}{\pi}\ln2B(a)+\frac{2\ln2}{\pi}B\big(\frac{L-a}{2}\big)-\{a\to b\}\right].\hspace{3cm}
\end{eqnarray}
After this removal process and by using the Eq.\,\eqref{BSS.Def.}, the only finite parts remained from Eq.\,\eqref{subtraction.Mixed.rc} are:
\begin{eqnarray}\label{Remain.finite.terms.}
   E^{(1)}_{\hbox{\tiny{$\mathcal{M}$,Cas.}}}&=&\lim_{L\to\infty}\lim_{b\to\infty}\big[E_\mathcal{A}-E_\mathcal{B}\big]\nonumber\\
   &=&\lim_{L\to\infty}\lim_{b\to\infty}\Bigg\{\frac{-\lambda}{32a}\left[\frac{2a\ln2}{\pi}B(a)+B^2(a)+\frac{a}{4m}\tanh\left(ma\right)\right]\nonumber\\
  &+&2\frac{-\lambda}{32\frac{L-a}{2}}\left[\frac{(L-a)\ln2}{\pi}B\Big(\frac{L-a}{2}\Big)+B^2\Big(\frac{L-a}{2}\Big)
   +\frac{L-a}{8m}\tanh\left(\frac{m(L-a)}{2}\right)\right]-\{a\to b\}\Bigg\}.
\end{eqnarray}
At the final step, all limits presented in Eq.\,\eqref{Remain.finite.terms.} should be applied. This limit leads to the values of all branch cut terms related to the regions $A2$, $B1$ and $B2$ diminishes. Therefore, the final expression for radiative correction to the Casimir energy of massive scalar field confined with mixed BC between two points with distance $a$ in $1+1$ dimensions becomes:
\begin{eqnarray}
\label{final.RC.Mixed.Cas.}
    E^{(1)}_{\hbox{\tiny{$\mathcal{M}$,Cas.}}}=\frac{-\lambda}{32a}\left[\frac{2a\ln2}{\pi}B(a)+B^2(a)
    +\frac{a}{4m}\left[\tanh\left(ma\right)-1\right]\right].
\end{eqnarray}
This result is convergent for any values of $a\neq 0$ and $m\neq 0$. Whereas, our result given in Eq.\,\eqref{final.RC.Mixed.Cas.} has similarity in the sign with the one reported in\,\cite{NN-DN-DD}\,(both are positive), however our answer differs from those of reported on that work. This difference may be caused by the type of counterterms employed in the renormalization program. The counterterms that we have used in this calculation, are consistent with the BC imposed. Whereas, the counterterms for this type of problem in the previous works are the \emph{free counterterm}. The calculation of the massless limit from above equation directly yield a divergent result. Thus, to obtain the finite answer for this particular limit, we go back to Eq.\,\eqref{Vacuum.A1.Mixed.RC} and put the mass of the field as zero. Therefore, we have:
\begin{eqnarray}
\label{massless.Mixed.1}
   E_{\hbox{\tiny{$\mathcal{M}$,$A1$}}}^{(1)}=\frac{-\lambda}{32a}\left[\left(\sum_{n=1}^{\infty}\frac{1}{\frac{(2n+1)\pi}{2a}}\right)^2
   +\frac{1}{2}\sum_{n=0}^{\infty}\frac{1}{(\frac{(2n+1)\pi}{2a})^2}\right]
\end{eqnarray}
Using Eq.\,\eqref{massless.Mixed.1} and the definition of the BSS presented in Eq.\,\eqref{BSS.Def.}, we obtain:
\begin{eqnarray}
\label{massless.Mixed.2}
   E_\mathcal{A}-E_\mathcal{B}=\frac{-\lambda}{32\pi^2}\left[a+2\frac{L-a}{2}-b-2\frac{L-b}{2}\right]
   \left[\left(\sum_{n=1}^{\infty}\frac{1}{(n+\frac{1}{2})}\right)^2
   +\frac{1}{2}\sum_{n=0}^{\infty}\frac{1}{(n+\frac{1}{2})^2}\right]=0.
\end{eqnarray}
This equation obviously shows that the Casimir energy for the massless scalar field confined with mixed BC between two points is exactly zero.
\par
\begin{figure}
    \hspace{0cm}\includegraphics[width=12cm]{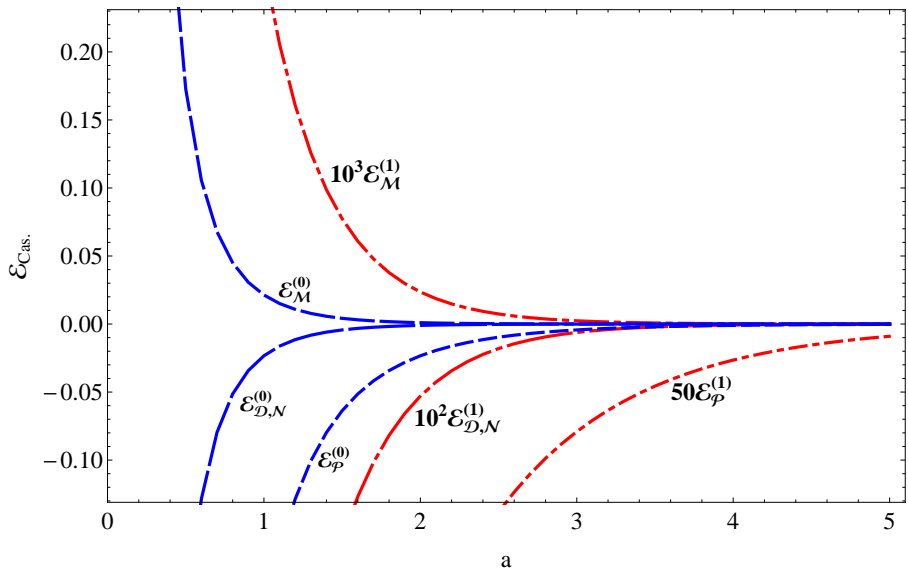}
 \caption{\label{All.Figs}   The zero and first-order radiative correction to the Casimir energy for massive scalar field confined between two points by distance $a$ with four Dirichlet, Neumann, Periodic and mixed BCs as a function of the distance of the points were plotted. The superscript $(0)$ denotes the zero (or leading) order term of the Casimir energy and the superscript $(1)$ denotes the first order term of the Casimir energy. In this figure, in addition to displaying the sign of the value of energies for each BC, we compare the magnitude of these energies with each other. The values of the mass and coupling constant in all plots are considered as $m=1$ and $\lambda=0.1$, respectively.}
\end{figure}

The relations between the expression of the Casimir energy for massive scalar field confined with Periodic, Neumann and Dirichlet BCs were reported as:
\begin{eqnarray}\label{Dirichlet.Neumann.Periodic.}
    E_{\hbox{\tiny{$\mathcal{P}$}}}(a)=2E_{\hbox{\tiny{$\mathcal{D}$}}}(a/2),\hspace{3cm}
    E_{\hbox{\tiny{$\mathcal{N}$}}}(a)=E_{\hbox{\tiny{$\mathcal{D}$}}}(a),
\end{eqnarray}
where $E_{\hbox{\tiny{$\mathcal{P}$}}}$, $E_{\hbox{\tiny{$\mathcal{D}$}}}$ and $E_{\hbox{\tiny{$\mathcal{N}$}}}$ denote the Casimir energy for the Periodic, Dirichlet and Neumann BCs between two points with distance $a$, respectively\,\cite{wolfram.}. It is worth noting that, the Casimir energy for massive scalar field with Dirichlet BC between two points was calculated in\,\cite{1D-Reza}. Thus, using Eq.\,\eqref{Dirichlet.Neumann.Periodic.} and the reported result for the Dirichlet Casimir energy in \,\cite{1D-Reza}, we can obtain the Casimir energy expression for Neumann and Periodic BCs. It can be conducted for both orders of the Casimir energy\,(zero and first-order) in massive and massless cases. In Fig.\,(\ref{All.Figs}), the values of zero and first-order radiative correction to the Casimir energy density with four types of BCs\,(Dirichlet, Neumann, mixed of them and Periodic) were plotted as a function of the distance of the points. In this figure, in addition to displaying the sign of the value of energies for each BC, one can also compare the magnitude of these energies with each other. Since, the Casimir energy value for Dirichlet, Neumann and Periodic BCs is negative, this value for the case of the mixed BC is positive for both orders of coupling constant $\lambda$.
\section{Conclusion}
\label{sec:conclusion}
In this article, the first-order radiative correction to the Casimir energy for massive and massless scalar fields confined with mixed BC between two points in $1+1$ dimensions were computed. An important point in this calculation is the use of a systematic method in renormalizing the bare parameters. Our used procedure allows all influences from the BCs to be imported in the renormalization program. One of the main results of importing the influences of the BC in the renormalization program is the appearance of position-dependent counterterms. Combining this renormalization program with the Box Subtraction Scheme\,(BSS) as a regularization technique, has resulted in a clear and unambiguous calculation process. In the BSS, the Casimir energy is usually defined by subtracting the vacuum energies from two similar configurations. The additional configurations add new parameters as a regulator in the computation process, and it leads to a clear and simple method in the removal process of the divergencies. These two procedures\,(the renormalization program and regularization technique), for the first time, were used to calculate the radiative correction to the Casimir energy for scalar field confined with the mixed BC, and the obtained answers were logically well-founded with a well-known physical basis. Note that, our obtained answer for first-order of the Casimir energy for massive scalar field is positive, and this value for the massless scalar field in the first-order of coupling constant $\lambda$ is obtained as zero.

\acknowledgments
The Author would like to thank the research office of Semnan Branch, Islamic Azad University for the financial support.

\end{document}